\documentclass[a4, 11pt]{article}
\usepackage{amsmath} %Never write a paper without using amsmath for its many new commands
\usepackage{amssymb} %Some extra symbols
\usepackage{geometry} % to change the page dimensions
\usepackage{graphicx} % support the \includegraphics command and options
\usepackage{epsfig}
\usepackage{booktabs} % for much better looking tables
\usepackage{array} % for better arrays (eg matrices) in maths
\usepackage{paralist} % very flexible & customisable lists (eg. enumerate/itemize, etc.)
\usepackage{verbatim} % adds environment for commenting out blocks of text & for better verbatim
\usepackage{subfig} % make it possible to include more than one captioned figure/table in a single float
\usepackage[dutch]{babel}
\usepackage{fancyhdr} % This should be set AFTER setting up the page geometry
\usepackage{sectsty}
\allsectionsfont{\sffamily\mdseries\upshape} % (See the fntguide.pdf for font help)
\usepackage[titles,subfigure]{tocloft} % Alter the style of the Table of Contents

\title{Towards Achieving Immersive Holographic-Type Communication}
\author{Filip De Turck, Ghent University - imec, Belgium}   
\date{}
\begin{document}
\maketitle
\renewcommand{\abstractname}{Abstract}
\renewcommand{\refname}{References}
\begin{abstract} 
This paper focuses on video streaming over telecommunication networks, taking the important evolution towards network softwarizaton into account. The importance and the opportunities provided by volumetric media delivery will be outlined by means of examples. The most appropriate management platform design for volumetric media delivery and the various challenges and possible approaches will be highlighted next. Finally an overview of research challenges and opportunities will be presented to generate further collaborative research in this area of research.
\end{abstract}

\section{Network resource management}
Network resource management is of prime importance for telecommunication network operators, equipment manufacturers and data center providers as it allows to (i)~make efficient use of the available resources, (ii)~offer service guarantees, and (iii)~make sure that services can be delivered with high quality of experience to end users.  Given the strong competition in the telecommunication domain and the increasing expectations of end-users, network operators and providers need reliable network resource management algorithms and methodologies. Ad hoc solutions often result in low resource utilization and high overhead, given the dynamic nature of resource availability and inherent complexity of resource allocation algorithms.  Examples of resource allocation algorithms are reported upon in~\cite{SupercomputingDeboosere} and~\cite{CNSM2017Santos}. Smart city use cases, where resource allocation is an important topic, are reported upon in~\cite{CNSM2017Santos, NOMS2018Santos}.

\section {Softwarized Networks} 
Softwarized networks bring virtualization concepts to the network and have proven to be particularly important for the industry, including telecommunication operators, cloud infrastructure and service providers. We refer to~\cite{vnfp, tnsmmoens, cnsm2021jose} for more information. The main advantages for network operators of softwarized networks is that they allow for (i)~replacement of dedicated hardware with generic hardware and software-based functions, (ii)~maximization of resource utilization and optimize energy usage, (iii)~faster and easier deployment, configuration, and updating of network functions, and (iv)~support for the Network-as-a-Service business model. The main advantages for service providers are that softwarized networks allow for (i)~dynamical scaling of the network, computing and storage resources based on the service requirements, and (ii)~reduced time to market for services.

\section{Adaptive service delivery}
Adaptive video delivery~\cite{scalablevideolaga, 6dofadaptive, reviewMariaQoE} is an important and widely deployed technology, where the video clients estimate the most appropriate video quality levels of the video segments and inform the server nodes about the quality level to select. Optimization of the delivery of adaptive video streaming services by in network optimizations~\cite{ToMBouten, ComNetBouten} is often considered by network and service providers. A framework for mobile augmented reality applications is published in~\cite{JSSVerbelen}.

\section{Volumetric media}

\subsection{Overview}
Volumetric Media is a technique that captures a three-dimensional space and often refers to technologies such Virtual Reality and Augmented Reality. Immersive means strong absorption into a technology for video streaming. Volumetric media are also often referred to as a holograms. In a hologram, parallax is added compared to a traditional 3D image. Parallax is the apparent displacement of an object because of a change in the observer's viewpoint. As a consequence, for hologram creation and consumption, 6 degrees of freedom (also referred to as 6DoF) are needed. In contrast, 2 dimensions are used for traditional video streaming, e.g. for currently widely deployed teleconference applications.

\subsection{Uses cases}
Important use cases for volumetric media delivery include (i)~holographic collaboration and conferencing~\cite{mmsys2018}, (ii)~tele-surgery and remote patient monitoring~\cite{applsciHeyse}, and (iii)~remote industrial monitoring and management~\cite{jnsmMaria}.

\subsection{Hologram BW requirements}
The bandwidth requirements for volumetric media delivery are described in~\cite{commagclemm} for HD, 360 degree (4K), 360 degree (16K), point cloud based video and light field based video. Based on these, we can conclude that dynamic point-cloud scenes require a significant amount of data, for instance 20 GB/s when 4 point cloud objects are displayed simultaneously. Typically holograms are generated through a local application, because the current bandwidth requirements (Gb/s) are too high for remote access. Advanced compression techniques will be needed to significantly reduce the bandwidth requirements. 

\subsection{Subjective quality study}
In order to evaluate the perception of users of volumetric media when the content is streamed using adaptive streaming, we performed a subjective study~\cite{qomex2020hooft} with 30 users. Three types of video content were composed by four different point cloud objects presented on a flat screen and streamed using different bandwidth and bitrate allocation configurations. The main observations are that end-user based or over-the-top optimizations are currently not sufficient for the end-user, because these techniques do not contribute to the low latency requirements. For these reasons, a cross-layer based end-to-end architecture for volumetric media delivery will be needed, as detailed further below in this paper. 

\section{EU Spirit project}
The EU Horizon Europe SPIRIT project~\cite{spirit} (2022-2025) focuses on designing a scalable platform for innovations on real-time immersive telepresence. In particular, the following use cases are considered:
\begin{itemize}
\item Real-time human-to-human interactions
\begin{itemize}
\item holographic conversations,
\end{itemize}
\item Real-time human-machine interactions
\begin{itemize}
\item Human-initiated: multi-site machine supervision, supported by autonomous mobile robots for intralogistics,
\item Machine-initiated: autonomous mobile robots alert humans to solve contextual problems.
\end{itemize}
\end{itemize}
The project organizes two open calls for third parties to evaluate the designed platform on the project testbed. Target application domains are: healthcare, retail, education, training, entertainment, manufacturing, and tourism.

\section{Cross-layer architecture}
In order to achieve a performant and high quality immersive volumetric delivery, optimizations are needed at three different levels: (i)~end-user optimizations, (ii)~over-the-top and transport optimizations and (iii)~novel network architectures. These three levels are detailed below.
\subsection{End-user optimizations}
Client-side optimizations~\cite{netsoft2020sam} include (i)~viewport prediction, (ii)~synchronization of feeds and (iii)~cybersickness assessment and avoidance strategies. Server-side optimizations~\cite{jetcaspark} include for instance (i)~3D tiling and (ii)~multiple representations encoding.

\subsection{Over-The-Top and Transport optimizations}
Three are two flavors of traditional streaming: (i)~adaptive streaming: mainly for on-demand streaming with no real-time encoding and (ii)~live streaming, where WebRTC and QUIC are often used for browser-based real time streaming. The following Over-The-Top and transport layer optimizations are useful for volumetric media streaming:
\begin{enumerate}
\item low latency transport (combining advantages of TCP and UDP) (HTTP over QUIC, for example),
\item more intelligent buffering techniques (window based)~\cite{jetcaspark},
\item smarter retransmission techniques (including HTTP push)~\cite{acmtommhooft},
\item adaptive partially reliable delivery of immersive media over QUIC-HTTP/3.
\end{enumerate}

\subsection{Novel network architectures}
Two main innovations on the network layer are expected, which are very beneficial for volumetric media delivery:
\begin{enumerate}
\item Service Function Chaining: in this approach, the application for volumetric streaming is built by implementing several VNFs (Virtual Network Functions) and executing them as a SFC (Service Function Chain)~\cite{cnsm2021jose}. Examples of VNFs are: (i)~software component for stream capturing and scene merging, (ii)~VNF for compression and encoding, (iii)~VNF for stream transport and caching, (iv)~VNF for view prediction and decoding, and (v)~VNF for rendering and QoE (Quality of Experience) monitoring. The individual VNFs are executed on the most appropriate location (cloud, core network, edge, or fog) based on dynamic placement algorithms, as for instance described in~\cite{cnsm2021jose}. Other examples of placement algorithms are reported upon in~\cite{vnfp} and~\cite{ComComWauters}.
\item Distributed Network Architectures: in addition to SFC-based networking, networks will need to modernize to fully decentralized architectures where delays in some flows do not affect others. In Software Defined Networking, the current central solutions can evolve to new hierarchical and fully decentralized approaches, to enable lower flow setup times, as described in~\cite{nof2020hemanth}. Examples of hierarchical network management systems are reported upon in~\cite{hiermgmtsystem, hierMoens}.
\end{enumerate}

As an example framework, SRFog~\cite{cnsm2021jose} considers Fog Computing and micro-services for the VNF design. SRFog follows the Kubernetes architectural model and adopts container-based service chains and traffic flow optimization based on Segment Routing. The typical maximum end-to-end latency objective for immersive video delivery is 20 milliseconds. There is important ongoing work on taking latency and network bandwidth effectively into account in the Kubernetes container scheduling process, which is very beneficial for the management of volumetric media delivery streams.

\section{Conclusions and future work}
In order to be able to fulfill the quality expectations of users of future immersive media, there is a need for optimizations at all elements of the transmission chain as well as at all levels of the protocol stack. With the elements above, it is clear that we are on our way to achieving this. However, there is a still a lot of room for research. At the end-user fast encoding strategies, novel quality modelling as well as prediction and adaptive bitrate selection approaches will be needed to adapt the systems to the needs and requirements of the users. For the network domain, transport protocols will need to be made faster and more accurate. Moreover, network architectures will need to be updated taking decentralization and function placement into account.


\begin{thebibliography}{9}

\bibitem{vnfp} H. Moens, F. De Turck, VNF-P: A model for efficient placement of virtualized network functions, IEEE International Conference on Network and Service Management (CNSM), 2014, Rio De Janeiro, Brazil, pp. 418-423.

\bibitem{tnsmmoens} H. Moens, F. De Turck, Customizable function chains: Managing service chain variability in hybrid NFV networks, IEEE Transactions on Network and Service Management 13 (4), 2016, pp. 711-724.

\bibitem{scalablevideolaga} S. Laga, T. Van Cleemput, F. Van Raemdonck, F. Vanhoutte, N. Bouten, M. Claeys, F. De Turck, Optimizing scalable video delivery through OpenFlow layer-based routing, 2014 IEEE Network Operations and Management Symposium (NOMS), IEEE NOMS 2014, pp. 1-4.

\bibitem{commagclemm} A. Clemm, M. Torres Vega, H. K. Ravuri, T. Wauters and F. De Turck, Towards Truly Immersive Holographic-Type Communication: Challenges and Solutions, in IEEE Communications Magazine, vol. 58, no. 1, pp. 93-99, January 2020.

\bibitem{6dofadaptive} J. van der Hooft, T. Wauters, F. De Turck, C. Timmerer, H. Hellwagner, Towards 6dof http adaptive streaming through point cloud compression, Proceedings of the 27th ACM International Conference on Multimedia, ACM Multimedia, 2019, pp. 2405-2413.

\bibitem{qomex2020hooft} J. van der Hooft, M. Torres Vega, C. Timmerer, A. C. Begen, F. De Turck and R. Schatz, Objective and Subjective QoE Evaluation for Adaptive Point Cloud Streaming, 2020 Twelfth International Conference on Quality of Multimedia Experience (QoMEX), 2020, IEEE, ISBN: 9781728159652

\bibitem{reviewMariaQoE} M. Torres Vega, C. Perra, F. De Turck, A. Liotta, A review of predictive quality of experience management in video streaming services, IEEE Transactions on Broadcasting, Volume 64, Issue 2, IEEE, pp. 432-445.

\bibitem{spirit} SPIRIT: Scalable Platform for Innovations on Real-time Immersive Telepresence, EU Horizon Europe project, 2022-2025, https://www.spirit-project.eu/

\bibitem{mmsys2018} S. Gunkel, H. M. Stokking, M. J. Prins, N. van der Stap, F. B. ter Haar, and O. A. Niamut, Virtual reality conferencing: multi-user immersive VR experiences on the web, In Proceedings of the 9th ACM Multimedia Systems Conference (MMSys '18). Association for Computing Machinery, New York, NY, USA, 498–501. 

\bibitem{applsciHeyse} J. Heyse, M. Torres Vega, T. De Jonge, F. De Backere, and F. De Turck, A Personalised Emotion-Based Model for Relaxation in Virtual Reality, Journal Applied Sciences, MDPI, 2020, Volume 10, Number 17, Article Number 6124, 2076-3417.

\bibitem{jnsmMaria} M. Torres Vega, et al., Immersive Interconnected Virtual and Augmented Reality: A 5G and IoT Perspective, Journal of Network and System Management, JNSM, Springer, Volume 28, 2020, pp.. 796–826.

\bibitem{netsoft2020sam} S. Van Damme, M. Torres Vega and F. De Turck, Human-centric Quality Management of Immersive Multimedia Applications, 2020 6th IEEE Conference on Network Softwarization (NetSoft), 2020, pp. 57-64.

\bibitem{jetcaspark} J. Park, P. A. Chou, and J.-N. Hwang, Rate-Utility Optimized Streaming of Volumetric Media for Augmented Reality, IEEE JETCAS, vol. 9, 2018, pp. 149–62.

\bibitem{acmtommhooft} J. van der Hooft, M. Torres Vega, S. Petrangeli, T. Wauters, and F. De Turck, Tile-based Adaptive Streaming for Virtual Reality Video, ACM Transactions on Multimedia Computing, Communications, and Applications, Volume 15, Issue 4, November 2019, Article No. 110, pp 1–24.

\bibitem{SupercomputingDeboosere} L. Deboosere, B. Vankeirsbilck, P. Simoens, F. De Turck, B. Dhoedt, P Demeester, Efficient resource management for virtual desktop cloud computing, Springer Journal of Supercomputing 62 (2), 2012, pp. 741-767.

\bibitem{CNSM2017Santos} J. Santos, T. Wauters, B. Volckaert, F. De Turck, Resource provisioning for IoT application services in Smart Cities, IEEE International Conference on Network and Service Management (CNSM), 2017, Tokyo, Japan, pp. 1-9.

\bibitem{NOMS2018Santos} J. Santos, P. Leroux, T. Wauters, B. Volckaert, F. De Turck, Anomaly detection for smart city applications over 5g low power wide area networks, NOMS 2018-2018 IEEE/IFIP Network Operations and Management Symposium, IEEE NOMS 2018, pp. 1-9.

\bibitem{JSSVerbelen} T. Verbelen, T. Stevens, P. Simoens, F. De Turck, B. Dhoedt, Dynamic deployment and quality adaptation for mobile augmented reality applications, Journal of Systems and Software, Volume 84, Issue 11, 2011, Elsevier, pp. 1871-1882.

\bibitem{ComComWauters} T. Wauters, J. Coppens, F. De Turck, B. Dhoedt, P. Demeester, Replica placement in ring based content delivery networks, Elsevier Journal on Computer Communications, Volume 29, Issue 16, 2006, pp. 3313-3326.

\bibitem{ToMBouten} N. Bouten, S. Latr\'{e}, J. Famaey, W. Van Leekwijck, F. De Turck, In-network quality optimization for adaptive video streaming services, IEEE Transactions on Multimedia, 2014, Volume 16, Issue 8, pp. 2281-2293.

\bibitem{ComNetBouten} N. Bouten, R. de O Schmidt, J. Famaey, S. Latr\'{e}, A. Pras, F. De Turck, QoE-driven in-network optimization for Adaptive Video Streaming based on packet sampling measurements, Elsevier Computer Networks, Volume 81, 2015, pp. 96-115.

\bibitem{cnsm2021jose} J. Santos, J.van der Hooft, M. Torres Vega, T. Wauters, B. Volckaert, F. De Turck, Efficient orchestration of service chains in Fog Computing for immersive Virtual Reality, published in IEEE/IFIP/ACM CNSM 2021, Confererence on Network and Service Management, CNSM, 2021, pp. p.139-145.

\bibitem{nof2020hemanth} H. K. Ravuri, M. Torres Vega, J. van der Hooft, T. Wauters, B. Da and F. De Turck, "On Routing Scalability in Flat SDN Architectures," 2020 IEEE 11th International Conference on Network of the Future (NoF), 2020, pp. 23-27.

\bibitem{hiermgmtsystem} J. Famaey, S. Latr\'{e}, J. Strassner, F. De Turck, A hierarchical approach to autonomic network management, 2010 IEEE/IFIP Network Operations and Management Symposium Workshops, IEEE NOMS 2010, Osaka, Japan, pp 225-232.

\bibitem{hierMoens} H Moens, B Hanssens, B Dhoedt, F De Turck, Hierarchical network-aware placement of service oriented applications in clouds, IEEE Network Operations and Management Symposium, IEEE NOMS 2014, Cracow, Poland, pp. 1-8.

\end{thebibliography}
\end{document}